\documentclass[twoside,journal,web]{IEEEtran}
\usepackage{cite}
\usepackage{amsmath,amssymb,amsfonts}
\usepackage{algorithmic}
\usepackage{graphicx}
\usepackage{textcomp}
\usepackage{multirow}
\usepackage{setspace}
\usepackage{bigstrut}
\usepackage{array}
\usepackage{fancyhdr}

\begin{document}
\title{Diagnosis of Coronavirus Disease 2019 (COVID-19) with Structured Latent Multi-View Representation Learning}
\author{Hengyuan Kang$^\dagger$, Liming Xia$^\dagger$, Fuhua Yan$^\dagger$, Zhibin Wan, Feng Shi, Huan Yuan, Huiting Jiang, Dijia Wu,\\ He Sui, Changqing Zhang* and Dinggang Shen*\
\thanks{Manuscript received April 12, 2020; accepted April 26, 2020. This work was supported in part by National Natural Science Foundation of China (61976151 and 61732011), and the National Key Research and Development Program of China under Grant 2018YFC0116400. (H. Kang, L. Xia and F. Yan contributed equally to this work.) (Corresponding authors: Changqing Zhang and Dinggang Shen.)}
\thanks{H. Kang, Z. Wan and C. Zhang are with the College of Intelligence and Computing, Tianjin University, Tianjin 300350, China (e-mail: \{kanghengyuan, wanzhibin, zhangchangqing\}@tju.edu.cn).}
\thanks{L. Xia is with Department of Radiology, Tongji Hospital, Tongji Medical College, Huazhong University of Science and Technology, Wuhan, Hubei, China (e-mail:  xialiming2017@outlook.com).}
\thanks{F. Yan is with Department of Radiology, Ruijin Hospital, Shanghai Jiao Tong University School of Medicine, Shanghai, China (e-mail: yfh11655@rjh.com.cn).}
\thanks{F. Shi, H. Yuan, H. Jiang, D. Wu and D. Shen are with the Department of Research and Development, Shanghai United Imaging Intelligence Co., Ltd., Shanghai, China (e-mail: \{feng.shi, huan.yuan, huiting.jiang, dijia.wu\}@united-imaging.com, Dinggang.Shen@gmail.com).}
\thanks{H. Sui is with Department of Radiology, China-Japan Union Hospital of Jilin University, Changchun, China (e-mail: suihe910402@126.com).}
}
\IEEEpubid{\begin{tabular}[t]{c@{}c@{}}\copyright 2020 IEEE.  Personal use of this material is permitted.  Permission from IEEE must be obtained for all other uses, in any current or\\ future media, including reprinting/republishing this material for advertising or promotional purposes, creating new collective works,\\ for resale or redistribution to servers or lists, or reuse of any copyrighted component of this work in other works.\end{tabular}}
\maketitle

\begin{abstract}
Recently, the outbreak of Coronavirus Disease 2019 (COVID-19) has spread rapidly across the world. Due to the large number of infected patients and heavy labor for doctors, computer-aided diagnosis with machine learning algorithm is urgently needed, and could largely reduce the efforts of clinicians and accelerate the diagnosis process. Chest computed tomography (CT) has been recognized as an informative tool for diagnosis of the disease. In this study, we propose to conduct the diagnosis of COVID-19 with a series of features extracted from CT images. To fully explore multiple features describing CT images from different views, a unified latent representation is learned which can completely encode information from different aspects of features and is endowed with promising class structure for separability. Specifically, the completeness is guaranteed with a group of backward neural networks (each for one type of features), while by using class labels the representation is enforced to be compact within COVID-19/community-acquired pneumonia (CAP) and also a large margin is guaranteed between different types of pneumonia. In this way, our model can well avoid overfitting compared to the case of directly projecting high-dimensional features into classes. Extensive experimental results show that the proposed method outperforms all comparison methods, and rather stable performances are observed when varying the number of training data. 
\end{abstract}


\section{Introduction}
\label{sec:introduction}
\begin{figure}[ht]
	\center
	\includegraphics[width=3.6in,height = 3.9in]{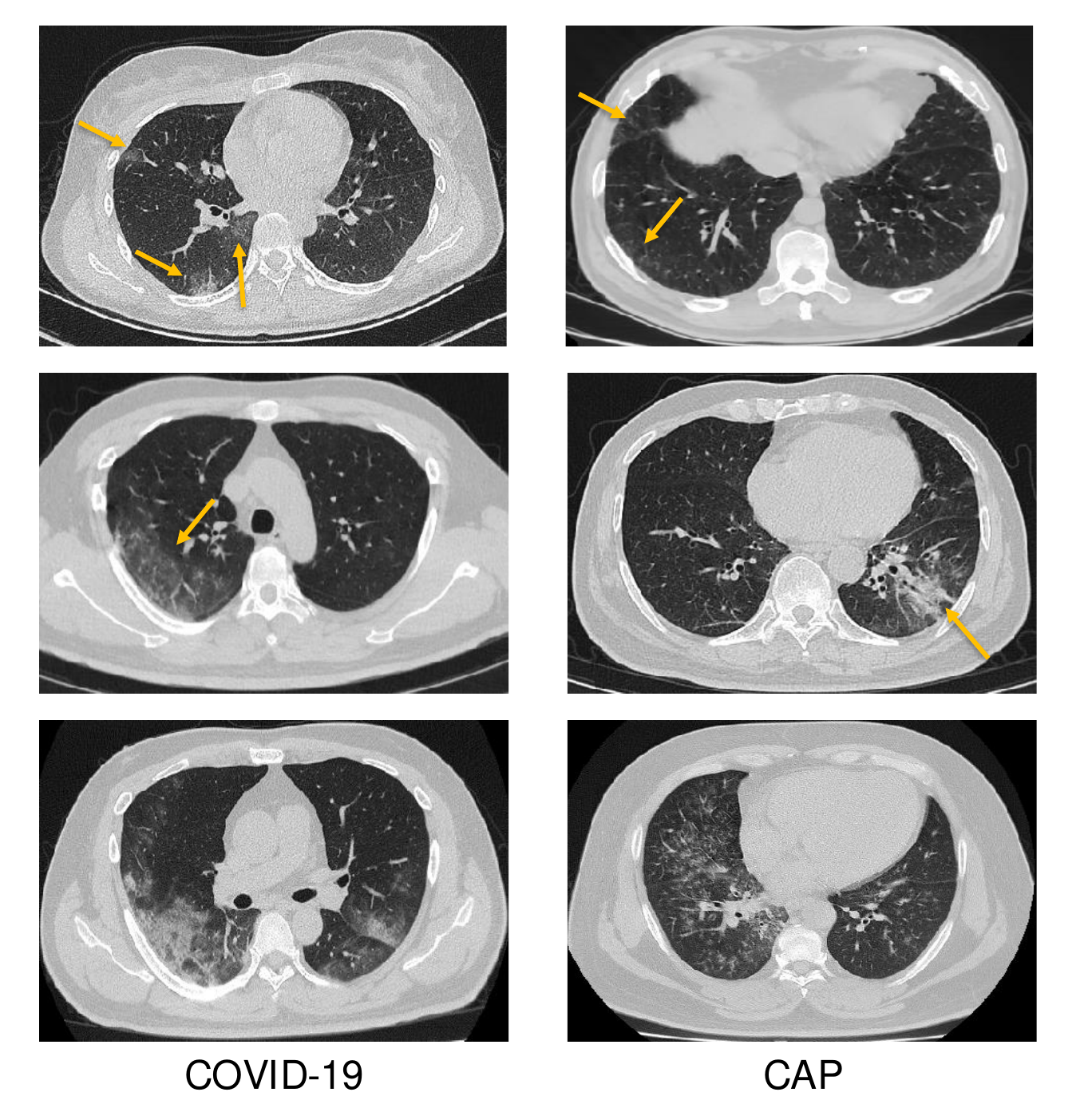}
	\caption{Examples of chest CT images with infection of COVID-19 (left) and community-acquired pneumonia (right). The pneumonia becomes more serious from top to bottom, and the yellow arrows indicate the representative infection areas. It can be observed that it is quite similar for the CT images of high-severity CAP and mild-severity COVID-19.}
	\label{fig:CT_example}
\end{figure}
\IEEEPARstart{A} novel coronavirus disease 2019 (COVID-19) was recognized in December 2019, in Wuhan, China and has rapidly spread over the world \cite{wu2020nowcasting,SHI2020Radiological,Song2020Emerging,world2020coronavirus,xu2020pathological}.
Recently, COVID-19 has been threatening all the world and the world health organization (WHO) has declared that COVID-19 becomes a global pandemic. 
The current clinical experience implies that the RT-PCR detection of viral RNA has a low sensitivity especially in the early stage \cite{li2020coronavirus,fang2020sensitivity,shi2020review,kanne2020chest}. As a form of pneumonia, inflammation of air sacs in lungs has been found, and it has shown that bilateral lung involvement could be observed for 
early, intermediate, and late stage patients. Accordingly, a high proportion of abnormal chest CT images were obtained from patients with this disease \cite{bernheim2020chest,kanne2020essentials,Zhao2020relation,chung2020ct}. Then, it is necessary to complement nucleic acid testing with automatic technique based on lung CT as one of the early diagnostic criteria for this new type of pneumonia as soon as possible.
In this study, 
\\
\\
we focus on conducting diagnosis for COVID-19 and community-acquired pneumonia \cite{zech2018variable,kermany2018identifying,rajpurkar2018deep}, i.e., characterizing the relationships between multiple types of features from CT images and these diseases, which provides a possible pipeline for automatic diagnosis and investigation. Specifically, multiple types of features are extracted and the correlations to diagnosis are extensively evaluated by conducting experiments with multiple baseline methods. The experiment results show that both radiomic features and handcrafted features are helpful for classifying these two different categories of lung diseases. Therefore, we propose a novel pipeline which can effectively integrate information from different views. We also note that, although deep-learning methods \cite{lecun2015deep},  especially CNN-based models, have shown the power in image classification, they usually need large-scale training data and have difficulty in exploiting expert prior. Although there are thousands of CT images available which is a quite large dataset for medical image analysis, it is still not comparable with the large-scale image dataset in the computer vision field, i.e., ImageNet \cite{deng2009imagenet}. Therefore, extracting manually designed features to incorporate expert prior is a reasonable and probably preferred approach which could alleviate the overfitting problem in machine learning. 

For computer-assisted medical diagnosis, a number of methods automatically learn classification models based on features extracted with expert prior \cite{mori2017computer,zhang2016automatic}. However, some of them are only applicable to single type of features thus cannot well explore the complementary information from multiple types of features. Fortunately, multi-view representation learning provides tools for exploiting multiple types of heterogeneous features \cite{liu2015inherent,zhang2017multi}. Although significant progress achieved, existing multi-view representation learning methods cannot guarantee the information completeness and promising structure for separability. This makes them tend to overfit training data thus harm the performance in the testing stage.

Based on the above analysis, we propose a new classification framework to diagnose these diseases, where the main aim is to identify the COVID-19 from CAP. It is worth noting that these different types of features are not directly used as input for a classifier to output the final decision \cite{wang2020deep,xu2020deep,gray2013random} in our method. Instead, both the training and testing samples are mapped into a promising latent space \cite{zhang2019cpm}, where these latent representations are expected to encode complementary information from different types of features and have promising structures revealing the underlying class distribution. 

First, since there are different types of features which are quite different in distribution and quality as shown in Fig.~\ref{fig:views}, it is very challenging and important to effectively integrate these different types of features. For this issue, a novel integration strategy is proposed with a group of neural networks, where each one encodes the information from one type of features into a latent representation. Second, we conduct projection learning to build an accurate model to map a subject with these multiple types of features into a latent representation, thus a latent-representation regressor is obtained which can be applied on new subjects. Third, a final classifier is trained based on the latent representation, instead of the original features. We should emphasize the advantages of our model over existing methods that often directly learn projections from original features into class labels. The first advantage is the latent representation, which is usually compact and thus may be more effective since it can avoid to overfit the high-dimensional data and has better generalization in the testing stage. Second, the proposed pipeline can encode information from different types of features and produce a structured representation which imposes simple bias on the model to further enhance the generalization performance. Moreover, the learned representation could be used in different classification models, and the performance with the learned latent representations clearly outperforms that of original features for all baseline classifiers used in experiments. The main contributions of this study are summarized as follows:
\begin{itemize}
\item We propose to conduct diagnosis of COVID-19 with multi-view representation learning, where the complementarity among different types of features is well explored, achieving clear performance gain in classification.
\item We propose a full pipeline for diagnosis COVID-19 from community-acquired pneumonia, which is quite different from existing models that directly project features into the class space. The key component is the structured latent representation learning which brings robustness, generalization and stability into the pipeline.
\item The learned latent representation can be widely used for different classifiers to promote the diagnosis accuracy. Specifically, the latent representations are adopted by several baseline models, and the results clearly demonstrate the effectiveness of the latent representation compared with original features.
\item Extensive experiments on the CT images validate that the proposed model can obtain a well-structured latent representation, thus significantly promoting the diagnosis in terms of accuracy, sensitivity and specificity compared with different methods.
\end{itemize}
\section{Material}
\label{sec:Dateset and Preprocessing}
\begin{figure}[htb]
	\center
	\includegraphics[width=3in,height = 2.1in]{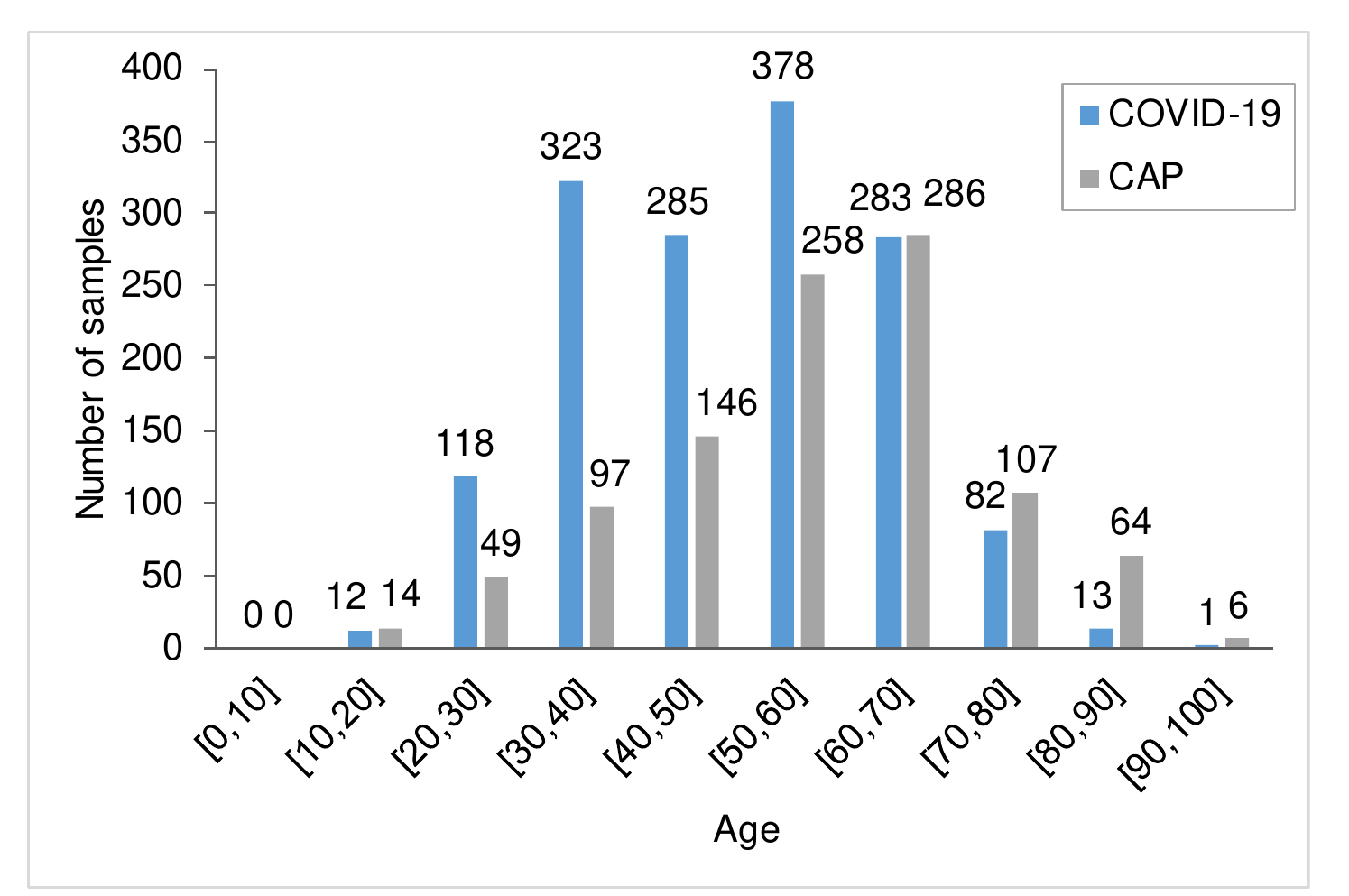}
	\caption{Subject distributions in terms of age.}
	\label{fig:age}
\end{figure}

There are 2522 CT images involved in this study, where 1495 cases are from COVID-19 patients and the left 1027 cases are from community-acquired pneumonia (CAP) patients. These COVID-19 infected subjects were confirmed with positive nucleic acid testing and confirmed by Chinese Centers for Disease Control and Prevention (CDC). The distributions of subjects in terms of gender and age are shown in Table~\ref{tab:gender} and Fig.~\ref{fig:age}. Specifically, according to Table \ref{tab:gender}, we can find that the number of males with COVID-19 is slightly larger than that of females, while,  for CAP, the contrary is the case. These subjects cover patients from 12-year old to 98-year old. In addition, basically, the average age of patients infected with COVID-19 is younger than that of CAP according to Fig.~\ref{fig:age}. These CT images were provided by Tongji Hospital of Huazhong University of Science and Technology, China-Japan Union Hospital of Jilin University, Ruijin Hospital of Shanghai Jiao Tong University, and their collaborators. The COVID-19 images were acquired from Jan. 9, 2020 to Feb. 14, 2020, and CAP images were obtained from Jul. 30, 2018 to Feb. 22, 2020.
\begin{table}[htb]
	\centering
	\caption{Subject distributions in terms of gender.}
	\begin{tabular}{ccc|c}
		\hline
		Gender & COVID-19 & CAP & Total \bigstrut\\
		\hline
		Male & 770 & 488 & 1258 \bigstrut[t]\\
		Female & 725 & 539 & 1264 \bigstrut[b]\\
		\hline
		Total & 1495 & 1027 & 2522 \bigstrut\\
		\hline
	\end{tabular}%
	\label{tab:gender}%
\end{table}%
\begin{figure*}[htb]
\center
\includegraphics[width=6.6in,height = 4.1in]{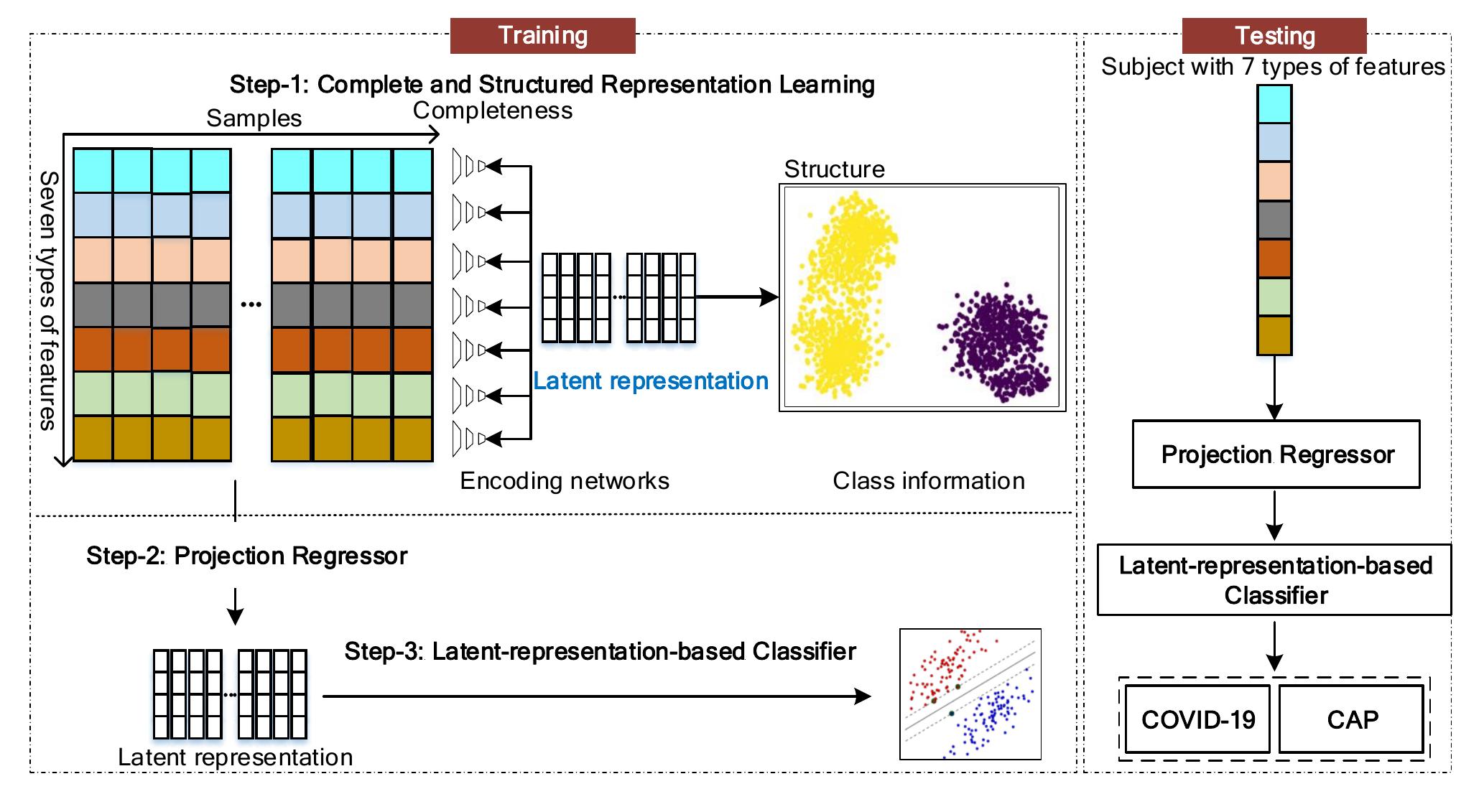}
\caption{Overview of the proposed latent-representation-based diagnosis framework. For the input, different colors indicate different types of features, while for the class information yellow and purple indicate COVID-19 and CAP, respectively.  }
\label{fig:framework}
\end{figure*}

Chest CT scans were performed on all patients with thin section. The CT scanners used in the three hospitals mentioned include uCT 780 from UIH, Optima CT520, Discovery CT750, LightSpeed 16 from GE, Aquilion ONE from Toshiba, SOMATOM Force from Siemens, and SCENARIA from Hitachi. Furthermore, CT scanners were carried out with the protocol which include: 120 kV, reconstructed CT thickness ranges from 0.625 to 2mm, with breath hold at full inspiration.

In this study, all images were preprocessed with a V-Net model\cite{milletari2016v} to extract the lung, lung lobes, and pulmonary segments. The infected lesions were also segmented\cite{shan+2020lung}. After that, based on the lesion region, 189-dimensional features were extracted from each CT image in total. First, according to different approaches of extracting, we divide the features into radiomic features and handcrafted features.
Furthermore, we split the radiomic features into two types of features that characterize the CT images from different perspectives:
\textbf{Gray features} are composed of the first-order statistics which describe the distribution of voxel intensities within the image region, such as maximum, minimum, median and so on. \textbf{Texture features} are derived from gray level co-occurrence matrix (GLCM), gray level size zone  matrix (GLSZM), gray level run length matrix (GLRLM), neighboring gray tone difference matrix (NGTDM) and gray level dependence matrix (GLDM) \cite{zwanenburg2016image}.  The handcrafted features are divided into five groups based on different characteristics of the lesions:  \textbf{Histogram features} are composed of frequency of intensity level in the infection area at 30 equal bins which are divided from the intensity value ( between -1350 and 150 \cite{shi2020largescale}) range of lung area image. \textbf{Number features} are composed of the total number of infected areas in the bilateral lungs, lung lobes, and pulmonary segments, respectively. \textbf{Intensity features} are composed of the mean and variance value of infection areas. \textbf{Volume features} are composed of volume of infected areas of the whole lung, each lobe and pulmonary segment, and the percentage of the infected areas of the whole lung and two lobes. \textbf{Surface features} are composed of the infection surface and the lung boundary surface. Furthermore, they contain the distance of each infection surface vertex to the nearest lung boundary surface, and are divided into 5 ranges (3, 6, 9, 12 and 15 voxels (voxel spacing is 1.5mm)). We also calculate the number of infection surface vertices to the lung wall in terms of each range of distance, the percentage of infection vertex number against the whole infection vertices in each range, and the percentage of infection vertex number. These 189 features in total are spit into 7 different non-overlapping groups as shown in Table~\ref{tab:features}.

\begin{table}[htb]
	\centering
	\caption{Abstract for different types of features.}
	\begin{tabular}{c|c|c}
		\hline
		\multicolumn{1}{c|}{} & Feature groups & \multicolumn{1}{c}{\# of features} \bigstrut\\
		\hline
		\multicolumn{1}{c|}{\multirow{2}[4]{*}{Radiomic feature}} & Gray features & 19 \bigstrut\\
		\cline{2-3}          & Texture features & 74 \bigstrut\\
		\hline
		\multicolumn{1}{c|}{\multirow{5}[10]{*}{Handcrafted feature}} & Histogram features & 30 \bigstrut\\
		\cline{2-3}          & Number features & 24 \bigstrut\\
		\cline{2-3}          &  Intensity features & 2 \bigstrut\\
		\cline{2-3}          & Surface features & 7 \bigstrut\\
		\cline{2-3}          & Volume features & 33 \bigstrut\\
		\hline
	\end{tabular}%
	\label{tab:features}%
\end{table}%
\section{Method}
There are different types of heterogeneous features from CT images which provide complementary information to diagnosis the COVID-19, hence we employ multi-view machine learning technique \cite{xu2013Multi-view,zhang2017flexible,Zhang2017CVPR} for our task. Inspired by our previous network (CPM-Nets) \cite{zhang2019cpm}, we further develop a novel diagnosis pipeline to classify COVID-19 and community-acquired pneumonia (CAP). Specifically, these diverse types of features extracted from CT images have extremely different properties, therefore it is unreasonable and ineffective to directly concatenate them without prepossessing or machine learning technique, and this is also validated in experiments. 

To effectively exploit these multiple types of features from CT images, we propose a latent-representation-based diagnosis pipeline, which is composed of three components in the training stage as shown in Fig.~\ref{fig:framework}. First, based on the CPM-Nets we learn latent representations with information completeness and promising class structure. The latent representations act as bridge of different components. This step is termed as Complete and Structured Representation Learning. Second, for the consistency of latent space \cite{zhang2019cpm} between training and testing, we train a projection model termed as Latent-representation Regressor between the 7 types of original features and the latent representations.  Third, a latent-representation-based classifier for diagnosis is trained. Accordingly, in the testing stage, the original features are projected into latent space with latent-representation regressor and then the final diagnosis result can be obtained with the latent-representation-based classifier.
\begin{figure*}[htb]
	\center
	\includegraphics[width=5in,height = 4in]{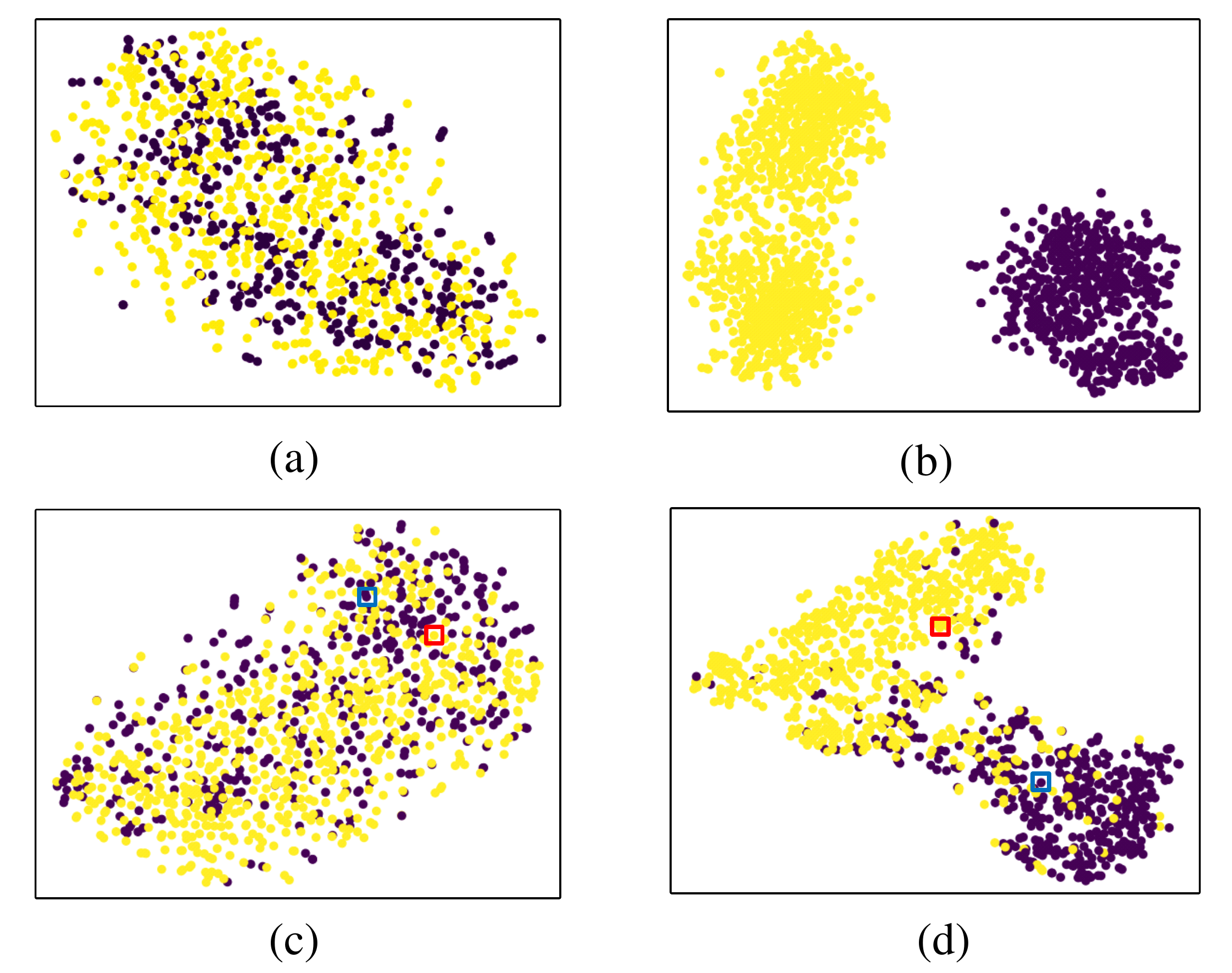}
	\caption{Visualization of the latent representations in the training and testing stages. Given the original features, the visualization in (a) indicates that the underlying class structure is not well revealed, while the learned latent representations in (b) are much better structured and consistent with classes. Similar case is also observed in the testing stage, as shown in (c) and (d). The red and blue boxes in (c) and (d) indicate the same pair examples for COVID-19 and CAP, respectively.}
	\label{fig:fitting}
\end{figure*}
\subsection{Step-1: Complete and Structured Representation Learning}
Considering our aim is to discriminate two types of CT images associated with COVID-19 and CAP, we learn latent representations which not only encode information of heterogeneous features but also reflect the class distribution. Then, the latent representations will be both informative and separable. For clarification, we first give the formal definition of notations as follows. Given the training set $\{\mathcal{X}_{n},y_{n}\}_{n=1}^{N}$, where $\mathcal{X}_{n} = \{\mathbf{x}_{n}^{(v)}\}_{v=1}^{V}$ is a multi-view sample and $y_{n}$ is the corresponding class label (i.e., $y_{n} = 1$ or $0$ indicates the subject is infected with COVID-2019 or CAP, respectively). $N$ and $V$ are the number of subjects in the training stage and the number of types of features (i.e., $V = 7$ in current experiments), respectively.
\subsubsection{Completeness for latent representation}
First, we aim to flexibly and effectively integrate different types of information for each subject into a low-dimensional space, where the desired latent representation should involve information from all types of features. From the perspective of reconstruction \cite{Imagerepresentationusing2DGaborwavelets}, if a latent representation $\mathbf{h}$ can well reconstruct each type of features with a stable mapping $f_{v}(\cdot)$, i.e., $\mathbf{x}^{(v)} =  f_{v}(\mathbf{h})$, then it encodes the intrinsic information of different types of features. Therefore, here we try to reconstruct each type of features from the learned latent representation to guarantee the information completeness.

Based on the above analysis, we can integrate information from different types of features into a latent representation as follows:
\begin{equation}
	\label{recoreconstruction loss}
	\ell_{r}(\mathcal{X}_{n},\mathbf{h}_{n}) = \sum_{v=1}^{V}\|f_{v}(\mathbf{h}_{n};\Theta_{r}^{(v)}) - \mathbf{x}_{n}^{(v)}\|^{2}
\end{equation}
where $f_{v}(\cdot;\Theta_{r}^{(v)})$ is the reconstruction network for the $v$th type of features parameterized by $\Theta_{r}^{(v)}$. $\mathbf{h}_{n}$ represents the learned latent representation. Ideally, by minimizing Eq.~(\ref{recoreconstruction loss}), we can well encode information from different types of features into the complete representation $\mathbf{h}_{n}$.
\subsubsection{Structure for latent representation}
Second, we aim to make the learned latent representation to be well structured with respect to these two different pneumonia diseases. Specifically, the loss for structured representations is specified as:
\begin{equation}
\Delta(y_{n},y) = \Delta(y_{n},g(\mathbf{h}_{n};\Theta_{c})),
\end{equation}
\begin{equation}
\text{with}~g(\mathbf{h}_{n};\Theta_{c}) = \mathop{\arg\min}_{y \epsilon \mathcal{Y}}\mathbb{E}_{\mathrm{h}\thicksim\mathcal{T}(y)}\mathit{F}(\mathbf{h},
\mathbf{h}_{n})
\end{equation}
where $\mathit{F}(\mathbf{h},\mathbf{h}_{n}) = \phi(\mathbf{h};\Theta_{c})^{T}\phi(\mathbf{h}_{n};\Theta_{c}) $ with $ \phi(\cdot;\Theta_{c}) $ being the feature mapping function for $\mathbf{h}$ parameterized by $ \Theta_{c} $, and $\mathcal{T}(y)$ being the set of latent representation from pneumonia class $y$. In practice, we set $\mathbf{h} = \phi(\mathbf{h};\Theta_{c})$ for simplicity which makes the loss non-parametric. This loss will enforce the compactness within the same type of pneumonia, and a margin is between COVID-19 and CAP, guaranteeing the separability. Then, we should minimize the following loss function
\begin{equation}
	\begin{aligned}
		\label{classificationloss}
		\ell_{c}(y_{n},y,\mathbf{h}_{n}) = \max_{y\epsilon \mathcal{Y}} \Big( 0,\Delta(y_{n},&y) + \mathbb{E}_{\mathbf{h}\thicksim\mathcal{T}(y)}\mathit{F}(\mathbf{h},
		\mathbf{h}_{n})\\
		&-\mathbb{E}_{\mathbf{h}\thicksim\mathcal{T}(y_{n})}\mathit{F}(\mathbf{h},\mathbf{h}_{n})\Big).
	\end{aligned}
\end{equation}
Based on above analysis, by jointly considering informativeness and separability, we should optimize the following objective function 
\begin{equation}
	\mathop{\arg\min}_{\Theta_{r}} \frac{1}{N}\sum_{n=1}^{N}\ell_{r}(\mathcal{X}_{n},\mathbf{h}_{n};\Theta_{r}) + \lambda\ell_{c}(y_{n},y,\mathbf{h}_{n}).
\end{equation}
where $\lambda > 0$ is a balance factor between completeness and class labels.

\subsection{Step-2: Learning Projection from Original Features to Latent Representation}
In the Step-1, one low-dimensional latent representation $\mathbf{h}$ is obtained for each subject in the training set. As shown in Fig.~\ref{fig:framework}, these representations are rather promising for the diagnosis of COVID-2019 and CAP since the representations associated with COVID-19 and CAP are compact and there is a clear margin between these two types of pneumonia. However, we should note that we cannot obtain the latent representation in the testing stage now. Therefore, we target to design a latent-representation regressor to accurately transform the original features of a subject into a latent representation.

The latent-representation regressor is implemented with fully connected neural networks to learn a mapping $\Gamma(\cdot)$, i.e., $\mathbf{h}^{'}_{n} = \Gamma(\mathcal{X}_{n};\Theta_{e})$ with parameter $\Theta_{e}$ from the original features, i.e., $\mathbf{x}_{n}$ to the corresponding latent representations. 
Specifically, the regressor is composed of four fully-connected layers and two sigmoid layers. The optimization objective is to minimize the MSE (Mean Squared Error) loss between the output of the regressor and the corresponding latent representations. It can be formulated as
\begin{equation}
	\mathop{\arg\min}_{\Theta_e}\frac{1}{N}\sum_{n=0}^{N}\Big(\mathbf{h}_{n} - \Gamma(\mathcal{X}_{n};\Theta_{e})\Big)^2.
\end{equation}
Then, given multiple types of original features of a CT image, the corresponding latent representation can be calculated. 

\subsection{Step-3: Latent-Representation-Based Classifier}
After obtaining the latent representation, we then target to train a latent-representation-based classifier which can diagnosis the subjects between COVID-2019 and CAP. For simplicity, we employ a neural network with three fully-connected layers as the latent-representation-based classifier. The widely used cross-entropy loss is employed in our classification tasks. Then, we should minimize the following loss function
\begin{equation}
    \begin{aligned}
        \ell = -\frac{1}{N}\sum_{n=0}^{N-1}\Big[ y_{n}&\log(\Phi_{\boldsymbol{\theta}}(\mathbf{h}_{n}^{'})) \\
        &+(1-y_{n})\log(1-\Phi_{\boldsymbol{\theta}}(\mathbf{h}_{n}^{'})) \Big] 
     \end{aligned}
\end{equation}
where $\Phi_{\theta}(\mathbf{h}_{n}^{'})$ indicates the prediction of pneumonia type from the classifier with $\mathbf{h}^{'}$ as the input.

\subsection{Testing Stage}
After training, a full pipeline for diagnosis of COVID-19 and CAP is available. The latent-representation regressor and latent-representation-based classifier play an essential role in the testing phase, as shown on the right side of Fig.~\ref{fig:framework}. Specifically, subjects with different types of features are first transformed into latent representation and then the diagnosis result can be obtained with the latent-representation-based classifier.
\section{Experiment and Result}
\subsection{Experimental Setting}
We conduct extensive experiments on the CT images data to evaluate the proposed pipeline. The dataset is randomly divided into 70\% and 30\% for training and testing, respectively. Furthermore, we adopt 5-fold cross-validation strategy on the training data to tune the parameter $\lambda$ from the set $\{ 0.1, 1,10,100 \}$. In practice, promising performance can be expected given a relatively large value for $\lambda$, which is fixed as 100 in our experiments.
\begin{table}[htbp]
	\centering
	\caption{Evaluation of Preprocessing of Features.}
	\begin{tabular}{p{2em}<{\centering}p{2em}<{\centering}ccc}
		\hline
		\textbf{Method} &       & \textbf{Original} & \textbf{Normalized} & \textbf{Standardized} \bigstrut\\
		\hline
		\multirow{3}[2]{*}{\textbf{LR}} & ACC   & 63.50$\pm$3.88  & 89.01$\pm$1.21  & \textbf{89.19$\pm$1.21 } \bigstrut[t]\\
		& SEN   & 1.00$\pm$0.00  & 90.64$\pm$1.04  & 90.66$\pm$2.08  \\
		& SPE   & 0.00$\pm$0.00  & 86.15$\pm$2.05  & 86.82$\pm$2.38  \bigstrut[b]\\
		\hline
		\multirow{3}[2]{*}{\textbf{SVM}} & ACC   & 61.26$\pm$5.21  & 86.40$\pm$2.73  & \textbf{89.40$\pm$1.21 } \bigstrut[t]\\
		& SEN   & 1.00$\pm$0.00  & 90.51$\pm$2.31  & 89.90$\pm$1.92  \\
		& SPE   & 0.00$\pm$0.00  & 74.52$\pm$4.12  & 88.59$\pm$2.23  \bigstrut[b]\\
		\hline
		\multirow{3}[2]{*}{\textbf{GNB}} & ACC   & 71.69$\pm$3.79  & 73.01$\pm$2.97  & \textbf{75.60$\pm$2.13 } \bigstrut[t]\\
		& SEN   & 96.42$\pm$1.98  & 89.32$\pm$2.14  & 85.22$\pm$2.75  \\
		& SPE   & 23.31$\pm$6.92  & 71.22$\pm$3.46  & 71.98$\pm$5.26  \bigstrut[b]\\
		\hline
		\multirow{3}[2]{*}{\textbf{KNN}} & ACC   & 64.62$\pm$3.24  & \textbf{87.10$\pm$2.03 } & 86.40$\pm$2.44  \bigstrut[t]\\
		& SEN   & 1.00$\pm$0.00  & 89.44$\pm$2.78  & 89.22$\pm$3.35  \\
		& SPE   & 0.00$\pm$0.00  & 78.33$\pm$3.36  & 76.00$\pm$4.75  \bigstrut[b]\\
		\hline
		\multirow{3}[2]{*}{\textbf{NN}} & ACC   & 70.43$\pm$3.23  &92.31$\pm$1.87  & \textbf{93.90$\pm$2.04}  \bigstrut[t]\\
		& SEN   & 98.21$\pm$0.56  & 91.76$\pm$1.52  & 94.60$\pm$2.16  \\
		& SPE   & 21.79$\pm$4.82  & 82.01$\pm$3.11  & 91.70$\pm$2.53 \bigstrut[b]\\
		\hline
	\end{tabular}%
	\label{tab:perprocess}%
\end{table}%

\begin{table*}[htb]
	\centering
	\caption{Diagnosis performance with baseline methods using different types of features.}
	\resizebox{\textwidth}{!}{
	\begin{tabular}{c|c|ccccccccc}
		\hline
		\multicolumn{1}{c}{\textbf{Method}} & \multicolumn{1}{c}{} &{\textbf{Radiomic}} & \textbf{Handcrafted} & \textbf{GF} & \textbf{TF} & \textbf{HF} & \textbf{NF} & \textbf{IF} & \textbf{SF} & \textbf{VF} \\
		\hline
		\multirow{3}[2]{*}{\textbf{SVM}} & \textbf{ACC(\%)} & 85.23$\pm$2.21  & 86.47$\pm$2.34  & 81.14$\pm$1.92  & 85.11$\pm$1.22  & 76.05$\pm$3.64  & 70.09$\pm$4.67  & 69.35$\pm$4.32  & 84.24$\pm$2.32  & 80.76$\pm$2.06 \\
		& \textbf{SEN(\%)} & 86.40$\pm$1.86  & 88.97$\pm$1.33  & 91.14$\pm$1.02  & 87.52$\pm$1.27  & 90.47$\pm$1.52  & 77.56$\pm$3.89  & 93.92$\pm$2.04  & 88.33$\pm$2.86  & 89.06$\pm$2.18  \\
		& \textbf{SPC(\%)} & 83.33$\pm$2.03  & 82.59$\pm$2.01  & 65.04$\pm$4.23  & 87.00$\pm$1.96  & 54.48$\pm$4.54  & 59.27$\pm$5.28  & 30.44$\pm$6.43  & 77.33$\pm$3.87  & 67.62$\pm$4.88  \\
		\hline
		\hline
		\multirow{3}[2]{*}{\textbf{LR}} & \textbf{ACC(\%)} & 87.33$\pm$2.18  & 89.19$\pm$2.22  & 81.56$\pm$2.67  & 84.91$\pm$3.02  & 69.83$\pm$3.98  & 69.45$\pm$4.32  & 70.94$\pm$3.77  & 81.56$\pm$2.57  & 78.21$\pm$2.41 \\
		& \textbf{SEN(\%)} & 93.41$\pm$1.19  & 88.21$\pm$2.31  & 91.54$\pm$1.67  & 89.21$\pm$1.92  & 92.72$\pm$1.02  & 71.64$\pm$3.01  & 95.08$\pm$1.17  & 86.76$\pm$2.51  & 86.70$\pm$2.33  \\
		& \textbf{SPC(\%)} & 77.33$\pm$3.63  & 90.77$\pm$1.22  & 65.53$\pm$4.02  & 77.83$\pm$4.33  & 33.33$\pm$5.81  & 65.84$\pm$4.55  & 27.22$\pm$6.32  & 72.58$\pm$3.08  & 66.06$\pm$2.32  \\
		\hline
		\hline
		\multirow{3}[2]{*}{\textbf{GNB}} & \textbf{ACC(\%)} & 79.30$\pm$2.84  & 75.04$\pm$2.44  & 72.30$\pm$3.28  & 73.50$\pm$3.88  & 71.30$\pm$3.52  & 64.00$\pm$3.85  & 68.70$\pm$4.02  & 81.20$\pm$2.98  & 66.60$\pm$4.23 \\
		& \textbf{SEN(\%)} & 83.39$\pm$2.14  & 83.44$\pm$2.46  & 88.23$\pm$1.64  & 76.78$\pm$2.65  & 83.32$\pm$2.33  & 71.82$\pm$3.17  & 92.33$\pm$2.01  & 89.93$\pm$1.87  & 75.43$\pm$2.89  \\
		& \textbf{SPC(\%)} & 70.66$\pm$3.32  & 70.02$\pm$3.21  & 77.72$\pm$2.66  & 80.86$\pm$2.15  & 61.12$\pm$3.98  & 46.48$\pm$5.61  & 23.88$\pm$7.33  & 70.25$\pm$2.83  & 74.34$\pm$2.21  \\
		\hline
		\hline
		\multirow{3}[2]{*}{\textbf{KNN}} & \textbf{ACC(\%)} & 82.30$\pm$3.06  & 86.50$\pm$2.55  & 81.20$\pm$3.12  & 82.60$\pm$2.89  & 75.70$\pm$3.67  & 67.90$\pm$4.27  & 62.50$\pm$3.84  & 80.40$\pm$2.75  & 79.30$\pm$3.64\\
		& \textbf{SEN(\%)} & 85.33$\pm$2.24  & 86.23$\pm$2.31  & 89.32$\pm$1.75  & 84.23$\pm$2.61  & 88.32$\pm$2.16  & 70.87$\pm$3.11  & 88.22$\pm$2.07  & 84.02$\pm$2.36  & 80.39$\pm$2.84\\
		& \textbf{SPC(\%)} & 76.82$\pm$2.83  & 72.33$\pm$3.11  & 78.23$\pm$3.12  & 83.22$\pm$2.11  & 60.32$\pm$4.38  & 50.75$\pm$3.61  & 43.65$\pm$3.32  & 73.89$\pm$2.91  & 67.33$\pm$4.22\\
		\hline
		\hline
		\multirow{3}[2]{*}{\textbf{NN}} & \textbf{ACC(\%)} & 87.60$\pm$1.77  & 89.41$\pm$1.21  & 81.62$\pm$1.42  & 87.23$\pm$1.96  & 78.33$\pm$2.13  & 70.96$\pm$2.53  & 70.81$\pm$2.12  & 84.33$\pm$1.72  & 84.91$\pm$1.36\\
		& \textbf{SEN(\%)} & 90.11$\pm$1.03  & 91.16$\pm$1.25  & 86.62$\pm$1.86  & 88.63$\pm$1.77  & 92.24$\pm$1.32  &72.55$\pm$2.85   & 95.42$\pm$1.02  & 87.62$\pm$1.27  & 88.55$\pm$1.83\\
		& \textbf{SPC(\%)} & 83.06$\pm$2.21  & 87.33$\pm$1.89  & 70.91$\pm$2.63  & 84.12$\pm$1.63  & 52.76$\pm$3.26  & 68.43$\pm$3.07  & 28.03$\pm$4.48  & 79.25$\pm$2.53  & 76.31$\pm$3.16\\
		\hline
		\multicolumn{11}{p{62.5em}}{Note: GF, TF, HF, NF, IF, SF, VF represent gray features, texture features, histogram features, number features, intensity features, surface features and volume features, respectively. LR, GNB and NN are shorts for Logistic-Regression, Gaussian-Naive-Bayes algorithm and Fully-Connected-Neural-Networks, respectively.} \\
		\hline
	\end{tabular}}%
	\label{tab:everyview}%
\end{table*}%
\textbf{Data preprocessing.} The original features extracted from CT images are of rather different scales. Accordingly, data preprocessing is necessary before using them as input for learning algorithm. There are several data preprocessing strategies, e.g., normalization and standardization. Specifically, the standardization for $K$ features is computed as:
\begin{equation}
\hat{x^{i}} = \frac{x^{i} - \mu^{i}}{\sigma^{i}},  \qquad  i = 1,2,...,K
\end{equation}
where $ \mu^{i} $ and $ \sigma^{i} $ are mean value and standard deviation of the feature $ x^{i} $, respectively. $ \hat{x^{i}} $ denotes the standardization feature of original feature $x^{i}$. The features of normalization are calculated as:
\begin{equation}
\hat{x^{i}} = \frac{x^{i} - x_{min}^{i}}{x_{max}^{i} - x_{min}^{i}} , \qquad i = 1,2,...,K
\end{equation}
where $x_{min}^{i}$ and $x_{max}^{i}$ are minimum and maximum values of the feature $x^{i}$ respectively. Accordingly, $\hat{x^{i}}$ indicates the normalization feature of original feature $x^{i}$.

We conduct experiments on several baseline models by concatenating all the original features, normalized features and standardized features, respectively. The effects of the data preprocessing for diagnosis are shown in Table~\ref{tab:perprocess}. Specifically, the performance of using the original features is relatively low, the main reason of which may be the large scale difference among different features. Fortunately, in terms of accuracy, both of these preprocessing methods obtain a significant improvement ($1.32\% \sim 25.69\%$) on all classification models. For clarification and comparison fairness, we employ the standardized data for all methods in the following experiments. We compare the proposed method with the following methods in the diagnosis task, including SVM, Logistic-Regression (LR), Gaussian-Naive-Bayes (GNB), K-Nearest-Neighbors (KNN), and Fully-Connected-Neural-Networks (NN). For all these methods, we repeat 10 times and report the mean and standard deviation performance. Diagnostic performance is evaluated in terms of accuracy (ACC), sensitivity (SEN) and specificity (SPC).

\subsection{Performance Evaluation}
\begin{figure*}[htb]
	\centering
	\includegraphics[width=7.4in,height = 4.2in]{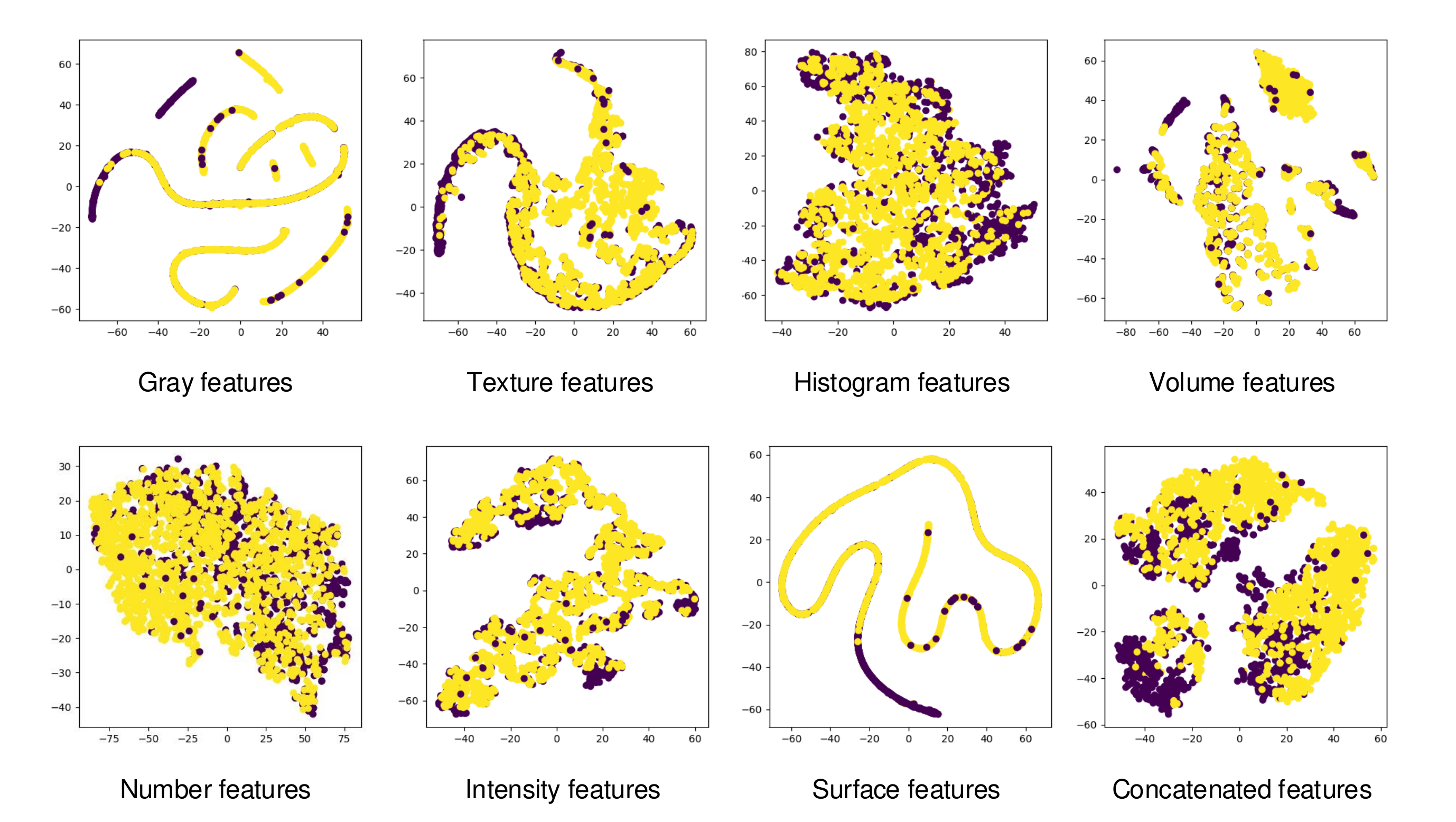}
	\caption{Visualization of each type of original features and concatenated features using  t-distributed stochastic neighbor embedding  (t-SNE)\cite{maaten2008visualizing},  which is a nonlinear dimensionality reduction technique well-suited for embedding high-dimensional data for visualization in a low-dimensional space.}
	\label{fig:views}
\end{figure*}

\subsubsection{Discrimination power of these different types of features}
In order to investigate the discrimination power (related to diagnosis) of different types of features, we visualize them with t-distributed stochastic neighbor embedding (t-SNE) \cite{maaten2008visualizing}. Fig.~\ref{fig:views} demonstrates different distributions for these 7 types of features and concatenated features (7 types). Furthermore, to quantitatively evaluate these features, we conduct experiments on each type of features for diagnosis task with baseline algorithms. Table~\ref{tab:everyview} presents the diagnostic performance. First, we can find that large performance gaps exist between different types of features.  For example, the baselines with gray features and texture features achieve clearly better performances 
than number features and intensity features. There are different manifestations reported between COVID-19 and other types of pneumonia, such as Influenza-A viral pneumonia \cite{xu2020deep}. As expected, radiomic features including gray and texture features have better discrimination ability. However, the number features and intensity features are a little less discriminative, and the possible reason is that the number of lesions and the intensity in lung may be not quite different for COVID-19 and CAP. Note that, although different types of features have different power in diagnosis, they are complementary to each other. As shown in Table~\ref{tab:everyview}, the concatenated features (i.e., radiomic and handcrafted features) perform much better than the case of using each individual type of features in terms of accuracy, which strongly supports the necessity of jointly using different types of features.

\begin{figure*}[]
	\center
	\includegraphics[width=7in,height = 1.5in]{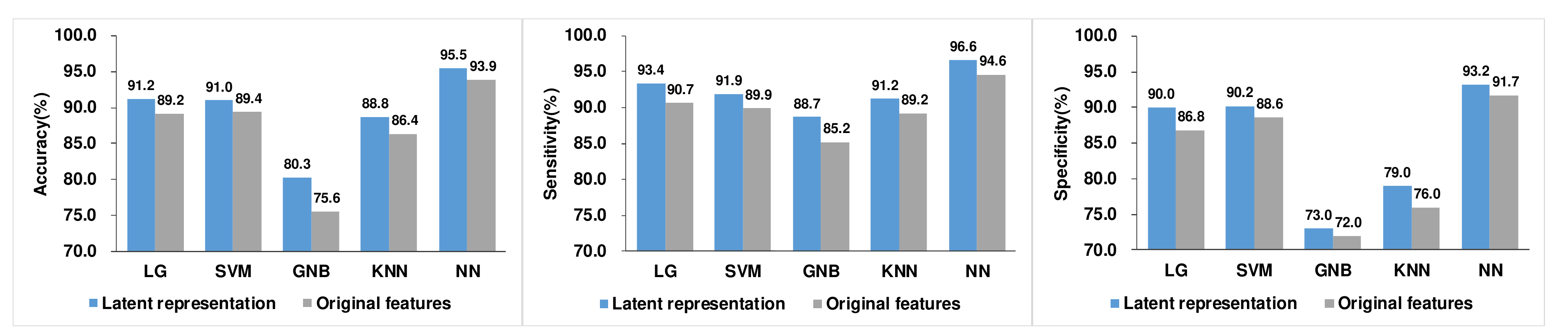}
	\caption{Evaluation for the latent representation on different classifiers.}
	\label{fig:Preformence with H}
\end{figure*}
\begin{figure}[htb]
	\center
	\includegraphics[width=3.2in,height = 2.2in]{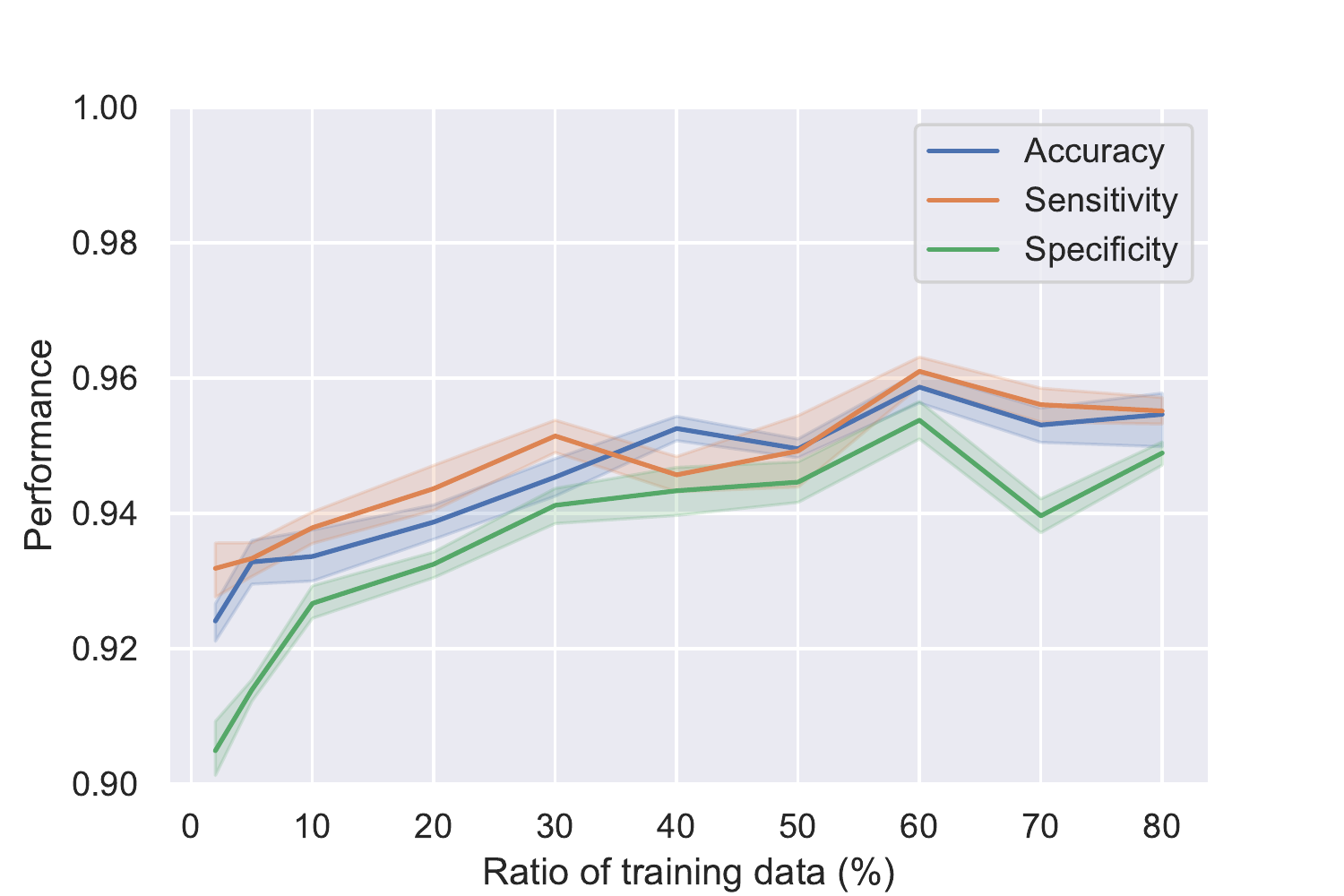}
	\caption{Stability of our method with different ratios of training data.}
	\label{fig:stable}
\end{figure}
\subsubsection{Effectiveness of latent representation compared with the original features}
With latent-representation regressor, Fig.~\ref{fig:fitting} intuitively demonstrates the effectiveness of the learned latent representation with informativeness and structure compared with the original features. Specifically, according to Fig.~\ref{fig:fitting}(a), it is observed that the original (concatenated) features are not well structured, while the learned latent representations (Fig.~\ref{fig:fitting}(b)) encoding information from original features and class labels can better reveal the underlying class structure. As expected, the counterparts Fig.~\ref{fig:fitting}(c) and Fig.~\ref{fig:fitting}(d) in the testing stage further validate the generalization ability of our model. Therefore, promising performance in diagnosis can be expected by using the learned latent representation.

For quantitatively evaluation, as shown in Fig.~\ref{fig:Preformence with H}, both conventional learning models and neural networks achieve significant improvement with the learned latent representation in terms of all three metrics. Specifically, Gaussian-Naive-Bayes obtains a clear performance improvement ($4.72\%$ and $3.44\%$) in terms of accuracy and sensitivity, respectively, while Logistic-Regression improves the performance by 3.13\% in terms of specificity. This further validates that the learned latent representation is effective in diagnosis for COVID-19 and CAP.

\subsubsection{Comparison for different methods}
Fig.~\ref{fig:Preformence with H} also shows diagnosis performance of the proposed method and compared methods. It is observed that the proposed method achieves the best accuracy, up to 95.50\%. Compared to all baselines which directly learn projections from the original features into class labels, our latent-representation-based model improves the diagnosis accuracy by $6.1\% \sim 19.9\%$. In terms of sensitivity and specificity, our method also demonstrates the best performance, improving the performance by $4.61\% \sim 21.22\%$ compared to the comparison methods.
To further investigate the effectiveness of our latent representation, we compare different diagnosis models by using the original features and our latent representation as shown in Fig.~\ref{fig:Preformence with H}. We can find that consistent better performances are achieved by using the latent representation for all classifiers. Furthermore, neural network with original features achieves promising performance, while the same structure neural networks using our latent representation achieve higher performance in terms of all metrics. This further validates the advantage and potential of the latent representation learned from our pipeline.

\subsubsection{Stability of proposed method}
We conduct experiments to verify the stability of our proposed model under different proportions of training data. For fair comparison, the testing set is fixed in each experiment. Fig \ref{fig:stable} reveals the fluctuation of performance as the ratio of training data changing from $2\%$ to $80\%$. We can find that the performance becomes clearly better as the number of training samples increases. However, when the ratio of training set exceeds $40\%$, the stability of performance could be observed, which also reflects the law of diminishing returns. For example, when $60\%$ of data are used, the model achieves the best results on all three metrics. While, the worst performance is only about $1\%$ lower than the best. Therefore, the promising performance and stable training results empirically validate that the proposed method can accurately and stably identify COVID-19 from CAP.
\section{Conclusion and Future Work}
In this study, we proposed a novel automatic diagnosis pipeline for COVID-19 which can fully leverage different types of features extracted from CT images. We investigated these different types of features and found they are complementary to each other. Then, with the proposed multi-view representation learning technique, diagnosis performance is promoted to $95.5\%$, $96.6\%$ and  $93.2\%$ in terms of accuracy, sensitivity and specificity, respectively. More importantly, compared to the original features, the learned latent representation has potential for utilization in different classifiers. In the future, we will consider diagnosis with more classes  (i.e., normal, different COVID-19 severity, and CAP) instead of only two types of disease (i.e., COVID-19 and CAP). Moreover, clinical characteristics for patients might be beneficial for diagnosis, which can be flexibly integrated into our framework for performance promotion.

\bibliographystyle{IEEEtran}
\bibliography{ref}

\end{document}